\documentclass[aps,prc,preprint,amsmath,amssymb,showpacs,preprintnumbers,superscriptaddress]{revtex4-1}

\usepackage[utf8]{inputenc}
\usepackage[sort&compress]{natbib}
\usepackage{ulem}
\usepackage{bm}
\usepackage{times}
\usepackage{amssymb,amsbsy,amsmath,amsfonts}
\usepackage{graphicx}
\usepackage{float}
\usepackage{color}
\usepackage{morefloats}
\usepackage{rotating}
\usepackage{srcltx}
\usepackage{slashed}
\usepackage{subfigure}
\usepackage{multirow}
\usepackage{verbatim}
\usepackage{hyperref}
\usepackage{tabularx}
\usepackage{braket}

\DeclareUnicodeCharacter{3000}{HEREHEREHERE}

\begin{document}

\title{Nuclear matter in relativistic Brueckner-Hartree-Fock theory with local and  nonlocal   covariant chiral interactions at leading order}




\author{Wei-Jiang Zou}
\affiliation{State Key Laboratory of Nuclear Physics and Technology, School of Physics, Peking University, Beijing 100871, China}
\author{Yi-Long Yang}
\affiliation{State Key Laboratory of Nuclear Physics and Technology, School of Physics, Peking University, Beijing 100871, China}
\author{Shihang Shen}
\email{sshen@buaa.edu.cn}
\affiliation{Peng Huanwu Collaborative Center for Research and Education, International Institute for Interdisciplinary and Frontiers, Beihang University, Beijing 100191, China}

\author{Jie Meng}
 \email{mengj@pku.edu.cn}
\affiliation{State Key Laboratory of Nuclear Physics and Technology, School of Physics, Peking University, Beijing 100871, China}
\affiliation{Center for Theoretical Physics, China Institute of Atomic Energy, Beijing, 102413, China}

\date{\today}

\begin{abstract}
The simultaneous description for nuclear matter and finite nuclei has been a long-standing challenge in nuclear \textit{ab initio} theory. 
With the success for nuclear matter, the  relativistic Brueckner-Hartree-Fock (RBHF) theory with covariant chiral interactions is a promising \textit{ab initio} approach to describe both nuclear matter and finite nuclei.
In the description of the  finite nuclei with the current RBHF theory, the covariant chiral interactions have to 
 be localized to make calculations feasible.
In order to examine the reliability and validity, in this letter, the RBHF theory with local and nonlocal covariant chiral interactions at leading order are applied for  nuclear matter.
The low-energy constants in the covariant chiral interactions determined with the local regularization are close to those with the nonlocal regularization. 
Moreover, the RBHF theory with local and nonlocal covariant chiral interactions provide equally well description of the saturation properties of nuclear matter. 
The present work paves the way for the implementation of covariant chiral interactions in RBHF theory  for finite nuclei.

\end{abstract}

\maketitle


\section{Introduction}

Nuclear matter is an idealized homogeneous system containing infinitely many neutrons and protons. 
Its equation of state~(EoS)~plays a vital role in our understanding of nuclear physics, such as the structure of finite nuclei, the reaction mechanisms of heavy-ion collision experiments and the evolution of neutron stars~\cite{Lattimer:2000kb,meng2016relativistic,Shen:2019dls,Sorensen:2023zkk}.
The saturation point of nuclear matter and its EoS in the vicinity serve as ideal benchmarks for testing the validity of nuclear forces and many-body theories.
Therefore, microscopic nuclear theories should aim to describe both the nuclear matter and the finite nuclei simultaneously.

However, it remains a nontrivial task to simultaneously describe the bulk properties of finite nuclei and the saturation properties of symmetric nuclear matter~(SNM) in \textit{ab initio} calculations starting from the realistic nucleon-nucleon~(NN)~interactions determined by free-space scattering.
In nonrelativistic \textit{ab initio} calculations, it is necessary to supplement the two-body forces with three-nucleon forces~(3NF)~ to reasonably describe the saturation of SNM~\cite{Li:2008bp,Hagen:2013nca,Ekstrom:2015rta,Sammarruca:2014zia,Drischler:2017wtt,Lonardoni:2019ypg,Sammarruca:2021bpn,Elhatisari:2022zrb}.
However, recent \textit{ab initio} calculations of nuclear structure using two- and three-nucleon interactions have revealed that the bulk properties of medium-mass nuclei and the saturation properties of SNM cannot be simultaneously described~\cite{Hoppe:2019uyw,Huther:2019ont,Machleidt:2023jws}.
The development of a consistent microscopic 3NF for finite nuclei and nuclear matter remains a challenge. 

Another distinct \textit{ab initio} approach involves treating finite nuclei and nuclear matter as relativistic systems within a unified theoretical framework. 
In Refs.~\cite{Anastasio:1980jm,Brockmann:1990cn}, in contrast to the nonrelativistic Brueckner-Hartree-Fock (BHF) theory, the relativistic Brueckner-Hartree-Fock~(RBHF)~approach demonstrates that saturation properties closer to the empirical values can be achieved using NN interactions alone. 
In recent years, this line of research has gained significant momentum following the overcoming of several long-standing technical problems~\cite{Brown:1985gt,Ma:2002fm,vanDalen:2004pn,Tong:2018qwx,Wang:2021mvg}. 
In particular, covariant chiral interactions have recently been developed, demonstrating their advantage in describing NN phase shifts~\cite{Ren:2016jna,Lu:2021gsb}.
Based on leading-order (LO) covariant chiral interactions, RBHF calculations have successfully described the saturation properties of SNM~\cite{Zou:2023quo}. 
This contrasts with the nonrelativistic BHF theory, where the use of only two-body interactions fails to reasonably describe the saturation of nuclear matter~\cite{Sammarruca:2021bpn}. 
With the success for nuclear matter, the RBHF theory with covariant chiral interactions is a promising \textit{ab initio} approach to describe both nuclear matter and finite nuclei .

The RBHF theory for finite nuclei~\cite{Shen:2016bva} is currently formulated in the coordinate space. 
Therefore, to implement the covariant chiral interactions, one needs to Fourier transform the interactions from momentum space to coordinate space.
However, in Ref.~\cite{Zou:2023quo}, a nonlocal regularization for the covariant chiral interactions is employed to avoid ultraviolet divergences.
In coordinate space, the nonlocality of interactions introduces additional complexities in the current RBHF theory for finite nuclei.
Therefore, the covariant chiral interactions must be locally regularized in order to make RBHF calculations for finite nuclei feasible.

In this Letter, we employ the LO covariant chiral interactions with the local and nonlocal regularization to study the properties of nuclear matter in the RBHF theory, to provide a first test for the reliability of local covariant chiral interactions.
We find that, the RBHF theory with local and nonlocal covariant chiral interactions provides an equally good description of the
saturation properties of nuclear matter.

\section{Theoretical framework}
The present relativistic \textit{ab initio} calculations are based on the covariant formulation of chiral effective field theory~(EFT)~\cite{Ren:2016jna}. 
The form of the LO covariant chiral interactions is given by~\cite{Ren:2016jna}
\begin{eqnarray}
    V_{\mathrm{LO}} = V_{\mathrm{CT}} + V_{\mathrm{OPE}}.
\end{eqnarray}
It comprises contact terms~(CT)~and a one-pion-exchange~(OPE)~term, where the contact terms are
\begin{eqnarray}
    V_{\mathrm{CT}}(\bm{p}',\bm{p}) 
    &=& C_S\bar{u}(\bm{p}')u(\bm{p})
    \bar{u}(-\bm{p}')u(-\bm{p})\nonumber\\
    &+& C_V\bar{u} (\bm{p}')\gamma_\mu u(\bm{p})
    \bar{u}(-\bm{p}')\gamma^\mu u(-\bm{p})\nonumber\\
    &+&C_{AV}\bar{u} (\bm{p}')\gamma_\mu\gamma_5 u(\bm{p})
    \bar{u}(-\bm{p}')\gamma^\mu\gamma_5 u(-\bm{p})\nonumber\\
    &+&C_T \bar{u} (\bm{p}')\sigma_{\mu\nu} u(\bm{p})
    \bar{u}(-\bm{p}')\sigma^{\mu\nu} u(-\bm{p}),
\end{eqnarray}
with $\bm{p}~(\bm{p}')$~the spatial component of initial (final) momentum of the nucleon in the center-of-mass frame. Here, the helicity and isospin indices are suppressed for simplicity. where~$C_{S,V,AV,T}$~are low-energy constants~(LECs)~determined by fitting to the NN scattering phase shifts. 

To avoid ultraviolet divergences and facilitate numerical calculations, the relativistic chiral potential $V_{\rm LO}$ is regularized
by the following Gaussian regulator
\begin{align}
f(\bm{p},\bm{p}')=\exp[-\bm q^2/\Lambda^2].\label{equ14}
\end{align}
The regularization function only depends on the transfer momentum $\bm q=\bm{p}'-\bm{p}$ and, therefore, is local in coordinate space.
As such, it is better suited to the current RBHF theory for finite nuclei, which is formulated in coordinate space~\cite{Shen:2019dls}.
It was first employed by the Nijmegen group~\cite{Nagels:1977ze} and later applied in the study of chiral interactions~\cite{Ordonez:1993tn}.

In the Brueckner theory, the effective interaction~($G$-matrix)~is obtained by summing ladder diagrams of the  realistic NN interaction while incorporating the Pauli principle. In the RBHF framework, the~$G$-matrix is obtained by solving the in-medium relativistic scattering equation. One of the most widely used scattering equations in the RBHF theory is the Thompson equation~\cite{Thompson:1970wt}, which is a relativistic three-dimensional reduction of the Bethe-Salpeter equation~\cite{Salpeter:1951sz}. The Thompson equation including medium effects in the rest frame of nuclear matter reads~\cite{Brockmann:1990cn},
\begin{eqnarray}
    G(\bm{p}',\bm{p}|\bm{P},W)
    =&& V(\bm{p}',\bm{p}|\bm{P})
    +\int\frac{\mathrm{d}^3k}{(2\pi)^3}V(\bm{p}',\bm{k}|\bm{P})\nonumber\\
    &&\times\frac{M^{\ast2}}{E_{\bm{P}+\bm{k}}^\ast E_{\bm{P}-\bm{k}}^\ast}
    \frac{Q(\bm{k},\bm{P})}{W-E_{\bm{P}+\bm{k}}-E_{\bm{P}-\bm{k}}}\nonumber\\
    &&\times G(\bm{k},\bm{p}|\bm{P},W)\label{equ9},
\end{eqnarray}
where~$\bm{P}=\frac{1}{2}(\bm{k}_1+\bm{k}_2)$~is the center-of-mass momentum, and~$\bm{k}=\frac{1}{2}(\bm{k}_1-\bm{k}_2)$~is the relative momentum of the two interacting nucleons with momenta~$\bm{k}_1$~and~$\bm{k}_2$, and ~$\bm{p}$,~$\bm{p}'$,~and~$\bm{k}$~are the initial, final, and intermediate relative momenta of the two nucleons scattering in nuclear matter, respectively. $W=E_{\bm{P}+\bm{p}}+E_{\bm{P}-\bm{p}}$~is the starting energy. $M^\ast$~and~$E^\ast_{\bm{P}\pm\bm{k}}$~are the effective masses and energies~[see Eq.~\eqref{equ8}]. The Pauli operator~$Q(\bm{k},\bm{P})$~only allows the scattering of nucleons to unoccupied states, i.e.,
\begin{eqnarray}
    Q(\bm{k},\bm{P})
    =\begin{cases}
    1,\quad |\bm{P}+\bm{k}|,|\bm{P}-\bm{k}|>k_F\\
    0,\quad\text{otherwise}
    \end{cases},
\end{eqnarray}
where~$k_F$~is the Fermi momentum. 

In the RBHF calculations, the single-particle motion of
nucleons in nuclear matter is described by the Dirac equation
\begin{eqnarray}
(\bm{\alpha}\cdot\bm{p}+\beta M+\beta\mathcal{U})u(\bm{p},\lambda)
=E_{\bm{p}}u(\bm{p},\lambda),\label{equ4}
\end{eqnarray}
where~$\bm{\alpha}$~and~$\beta$~are the Dirac matrices, $u(\bm{p},\lambda)$~is the Dirac spinor with momentum~$\bm{p}$, single-particle energy~$E_{\bm{p}}$~and helicity~$\lambda$, $M$~is the mass of the free nucleon, and~$\mathcal{U}$~is the single-particle potential operator, providing the primary medium effects. Due to time-reversal invariance, the spacelike component of the vector fields can be neglected, and as a result, the single-particle potential operator can be  expressed as~\cite{Brockmann:1990cn,Tong:2018qwx}: 
\begin{eqnarray}
    \mathcal{U} = U_S + \gamma^0U_0.
\end{eqnarray}
The momentum dependence of the scalar field~($U_S$)~and the timelike component of the vector fields~($U_0$)~is weak~\cite{Gross-Boelting:1998xsk,Wang:2021mvg} and therefore neglected. By introducing the following effective quantities, the Dirac equation in a medium can be expressed in a form analogous to the free Dirac equation.
\begin{eqnarray}
    M^\ast = M + U_S,\quad E^\ast_{\bm{p}} = E_{\bm{p}} - U_0,\label{equ8}
\end{eqnarray}
whose solution reads
\begin{eqnarray}
    u(\bm{p},\lambda) = \sqrt{\frac{E^\ast_{\bm{p}}+M^\ast}{2M^\ast}}
    \begin{pmatrix}1\\ \dfrac{2\lambda p}{E^\ast_{\bm{p}}+M^\ast}\end{pmatrix}\chi_{\lambda},
\end{eqnarray}
where~$\chi_{\lambda}$~is the Pauli spinor helicity basis. The covariant normalization is $\bar{u}(\bm{p},\lambda)u(\bm{p},\lambda)=1$.

Once $U_S$~and~$U_0$~of the single-particle potential operator are determined, the in-medium Dirac equation can be solved analytically. To achieve this, the matrix element of~$\mathcal{U}$ is constructed following Refs.~\cite{Anas83,Brockmann:1990cn} as
\begin{eqnarray}
    \Sigma(p)
    =\bar{u}(\bm{p},1/2)\mathcal{U}u(\bm{p},1/2)=U_S+\frac{E_{\bm{p}}^\ast}{M^\ast}U_0,\label{equ6}
\end{eqnarray}
where the direction of~$\bm{p}$~is taken along the~$z$~axis. Once~$\Sigma$~is obtained, $U_S$~and~$U_0$~can be determined via the following relations,
\begin{subequations}\label{equ7}
\begin{eqnarray}
\Sigma(p_1) = U_S + \frac{E_{p_1}^\ast}{M^\ast}U_0,\\
\Sigma(p_2) = U_S + \frac{E_{p_2}^\ast}{M^\ast}U_0,
\end{eqnarray}
\end{subequations}
where~$p_1=0.5k_F$~and~$p_2=0.7k_F$~describe two momenta that are in the Fermi sea.

On the other hand, the matrix elements of~$\mathcal{U}$~in Eq.~\eqref{equ6}~can be calculated
as the integrals of the antisymmetrized~$G$~matrix
\begin{eqnarray}
    \Sigma(p) = 
    \int_0^{k_F}\frac{\mathrm{d}^3p'}{(2\pi)^3}\frac{M^\ast}{E_{\bm{p}'}^\ast}
    \langle \bar{u}(\bm{p})
    \bar{u}(\bm{p}')
    |\bar{G}|u(\bm{p})u(\bm{p}')\rangle\label{equ8},
\end{eqnarray}
In the~$no\text{-}sea$~approximation~\cite{Serot:1997xg}, the integral is only performed for the single-particle states in the Fermi sea. 

Eqs.~\eqref{equ9},~\eqref{equ4},~\eqref{equ7}~and~\eqref{equ8}~constitute a set of coupled equations that need to be solved self-consistently;
See Ref.~\cite{Zou:2023quo} for the details.

\section{Results and discussions}
\begin{table}[h]
\caption{\label{tab:table4}Values of the LO LECs~(in units of $\mathrm{GeV}^{-2}$)~using local regularization function~with the cutoff ranging from 500 MeV to 700 MeV  and the corresponding $\tilde{\chi}^2$ per datum of NN scattering phase shift.}\label{Table1}
\begin{ruledtabular}
\begin{tabular}{cccccc}
$\Lambda(\mathrm{MeV})$&
$C_S$&
$C_V$&
$C_{AV}$&
$C_T$&
$\tilde{\chi}^2/N$\\
\hline
$500$-I &$-691.26$&$603.38$&$-187.96$&
$-87.56$&7.60\\
$500$-II &$-749.77$&$664.46$&$-217.41$&
$-101.72$&7.41\\
$600$-I  &$-672.28$&$616.35$&$-194.20$&
$-87.52$&4.18\\
$600$-II  &$-711.64$&$659.47$&$-216.41$&
$-98.15$&4.10\\
$700$-I  &$-656.98$&$614.94$&$-122.34$&
$-51.84$&3.59\\
$700$-II  &$-674.85$&$637.83$&$-145.61$&
$-62.93$&3.32\\
\end{tabular}
\end{ruledtabular}
\end{table}

The LECs with local regularization function are determined by performing a simultaneous fit to the~$J\leq1$~Nijmegen partial wave phase shifts of the~$np$~channel up to the laboratory kinetic energy~($E_{\mathrm{lab}}$)~of 100 MeV at six energy points~\cite{Stoks:1993tb}. In the fitting process, a global fit to the $np$ phase shifts including all the LECs with the $\chi^2$-like function
\begin{equation}
    \tilde{\chi}^2=\sum_i(\delta^i-\delta_{\rm PWA93}^i)^2,\label{equ13}
\end{equation}
where $\delta_i$ are theoretical phase shifts or mixing angles, and
$\delta_{\rm PWA93}^i$ are the PWA93 data. 
In addition, to study the impact of the cutoff on the results, the cutoff $\Lambda$ is varied from 500 MeV to 700 MeV.
The corresponding LECs~$C_{S,V,AV,T}$~using local regularization function are listed in Table \ref{Table1}. 
The LECs in covariant chiral interactions determined with local regularization here are close to those with nonlocal regularization in Ref.~\cite{Zou:2023quo}.  
For each momentum cutoff~$\Lambda$, we consider two sets of LECs at two different local minima in the parameter space that have a similar description of the phase shifts. Both sets are employed in the present nuclear matter calculations to estimate the systematic errors from the LECs.
Here, we calculate the nuclear saturation properties and EoSs of SNM and pure neutron matter (PNM), leaving the study of finite nuclei to a subsequent study.

\begin{figure}[h]
\includegraphics[scale=0.35]{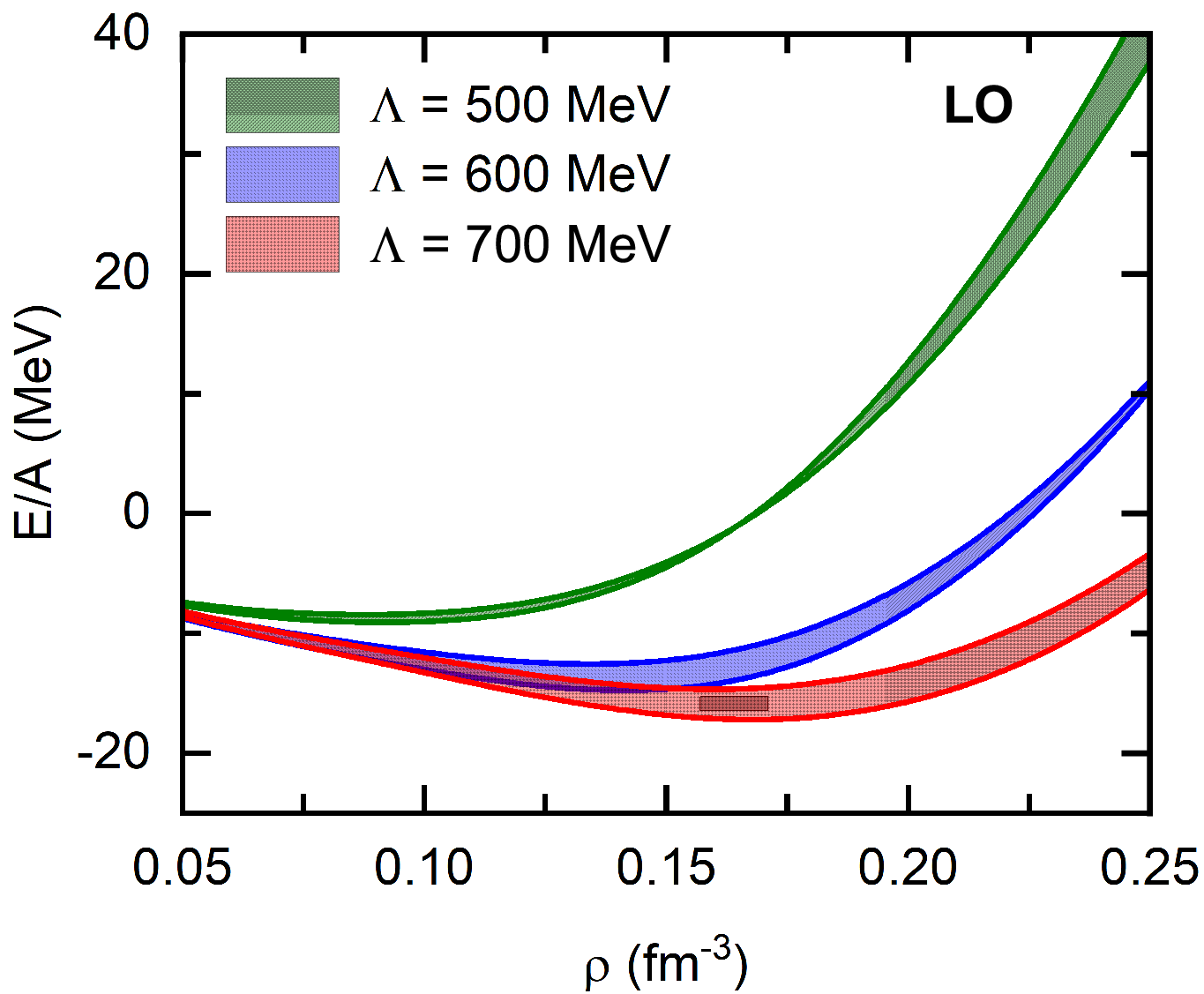}
\caption{\label{fig1}~(Color online)~Energy per nucleon~($E/A$)~in SNM as a function of the density~$\rho$~in the RBHF theory obtained with the LO covariant chiral interactions with a local regularization function.
The bands denote the spread of EoS obtained by two LEC sets at a given cutoff.
The shaded area indicates the empirical saturation region~\cite{Drischler:2017wtt}.}
\end{figure}

In Fig.~\ref{fig1}, we show the EoS for SNM calculated using the RBHF theory with the local LO covariant chiral interactions. The description of nuclear matter saturation is fairly good.
Overall, the EoSs become increasingly repulsive as the cutoff~$\Lambda$~decreases.
At low densities, we observe a weak cutoff dependence, and the nuclear matter saturation point can be reasonably described with momentum cutoff $\Lambda=600-700$ MeV.
In contrast, the energy per nucleon of SNM exhibits a stronger dependence on the momentum cutoff~$\Lambda$~at higher densities.
This is understandable since the chiral interactions are constructed based on a low-energy expansion.
We expect that the dependence of the results on the cutoff will be alleviated when higher-order covariant chiral interactions are employed.
For~$\Lambda=700$~MeV, the binding energy per nucleon obtained in the RBHF theory is in good agreement with the empirical value, and the saturation density is also in good agreement with the empirical value. 

\begin{figure}[h]
\includegraphics[scale=0.35]{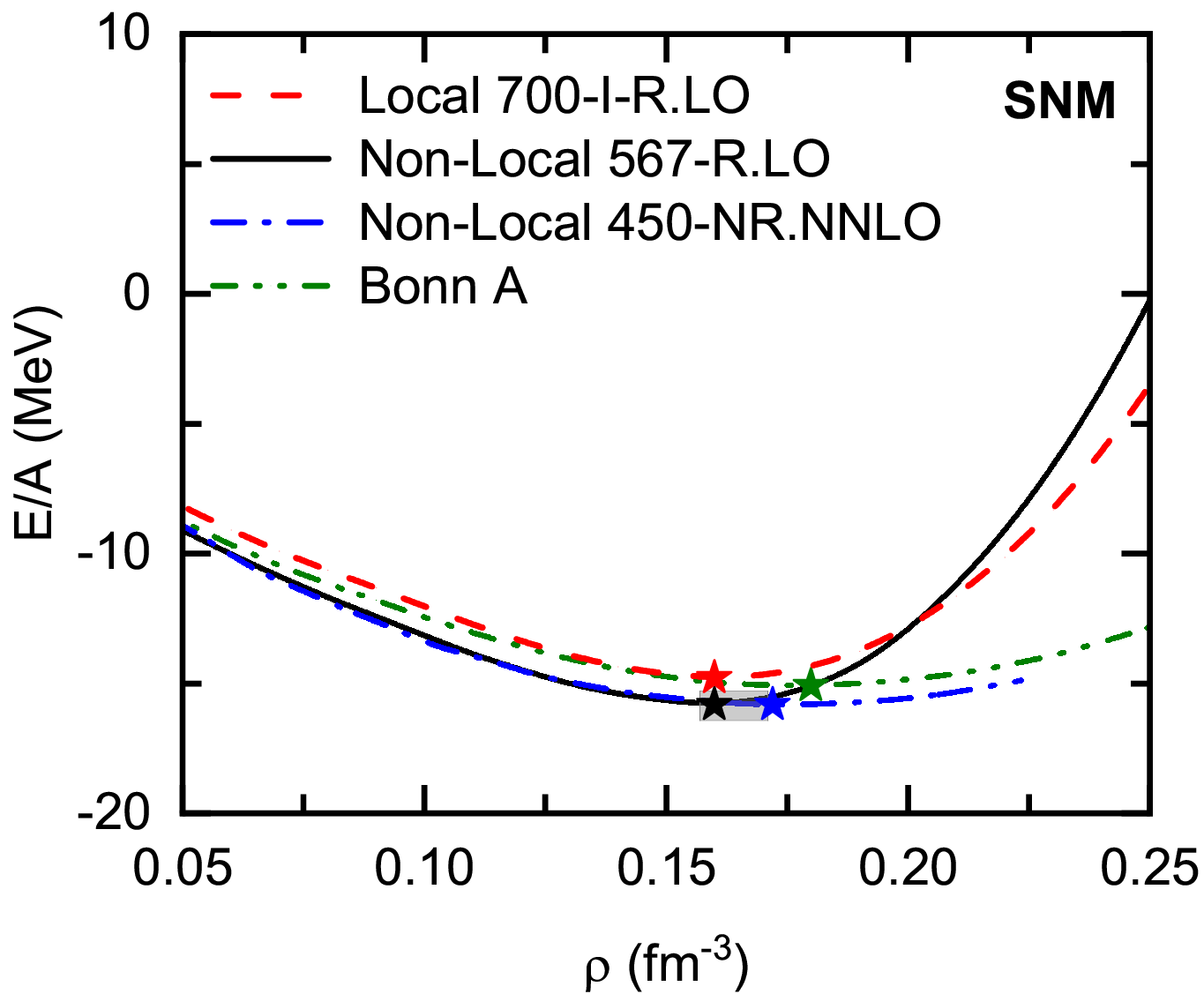}
\caption{\label{fig2}~(Color online)~Energy per nucleon~($E/A$)~in SNM as a function of the density~$\rho$~obtained from the LO covariant chiral interactions with local regularization~(red dash line), in comparison with the results from the LO covariant chiral interactions with nonlocal regularization~\cite{Sammarruca:2021mhv}~(black solid line), nonrelativistic NNLO chiral nuclear interactions~\cite{Sammarruca:2021bpn}~(blue dash-dot line) and Bonn A potential~\cite{Brockmann:1990cn}~(green dash-dot-dot line). The pentagrams denote the saturation point. The shaded area indicates the empirical values~\cite{Drischler:2017wtt}. The numbers in the legend indicate the momentum cutoff values.}
\end{figure}

In Fig.~\ref{fig2}, we compare the present results with the previous ones from the LO covariant chiral interactions employing a nonlocal regularization function~\cite{Zou:2023quo}, nonrelativistic next-to-next-to-leading-order~(NNLO)~calculations~\cite{Sammarruca:2021bpn}, and the Bonn A potential~\cite{Brockmann:1990cn}. 
For the LO covariant chiral interactions with local and nonlocal regularization, the LECs at cutoff $\Lambda=700$ MeV (LEC set 700-I in Table \ref{Table1}) and $\Lambda=567$ MeV are adopted, respectively.
Both the LO covariant chiral interactions with local and nonlocal regularization functions can reproduce the nuclear matter saturation point.
For the local (nonlocal) regularization, the calculated saturation point is $E/A=-14.79$~MeV and $\rho=0.16~\mathrm{fm}^{-3}$ ($E/A=-15.82$~MeV and $\rho=0.16~\mathrm{fm}^{-3}$).
Moreover, the EoSs of nuclear matter for both cases are quite similar.
They both show good agreement at low densities compared to the results from the Bonn A potential, but the EoSs become stiffer at densities beyond saturation density.
For the nonrelativistic chiral interactions, the nuclear saturation point can be reasonably described only when calculations are carried out up to NNLO with 3NF included.
In contrast, the RBHF theory with the LO covariant chiral interactions is sufficient to provide a decent description of the saturation properties of SNM, which indicates that nuclear matter saturation can indeed be understood as a relativistic effect~\cite{Anastasio:1980jm,Brockmann:1990cn}.

\begin{figure}[h]
\includegraphics[scale=0.35]{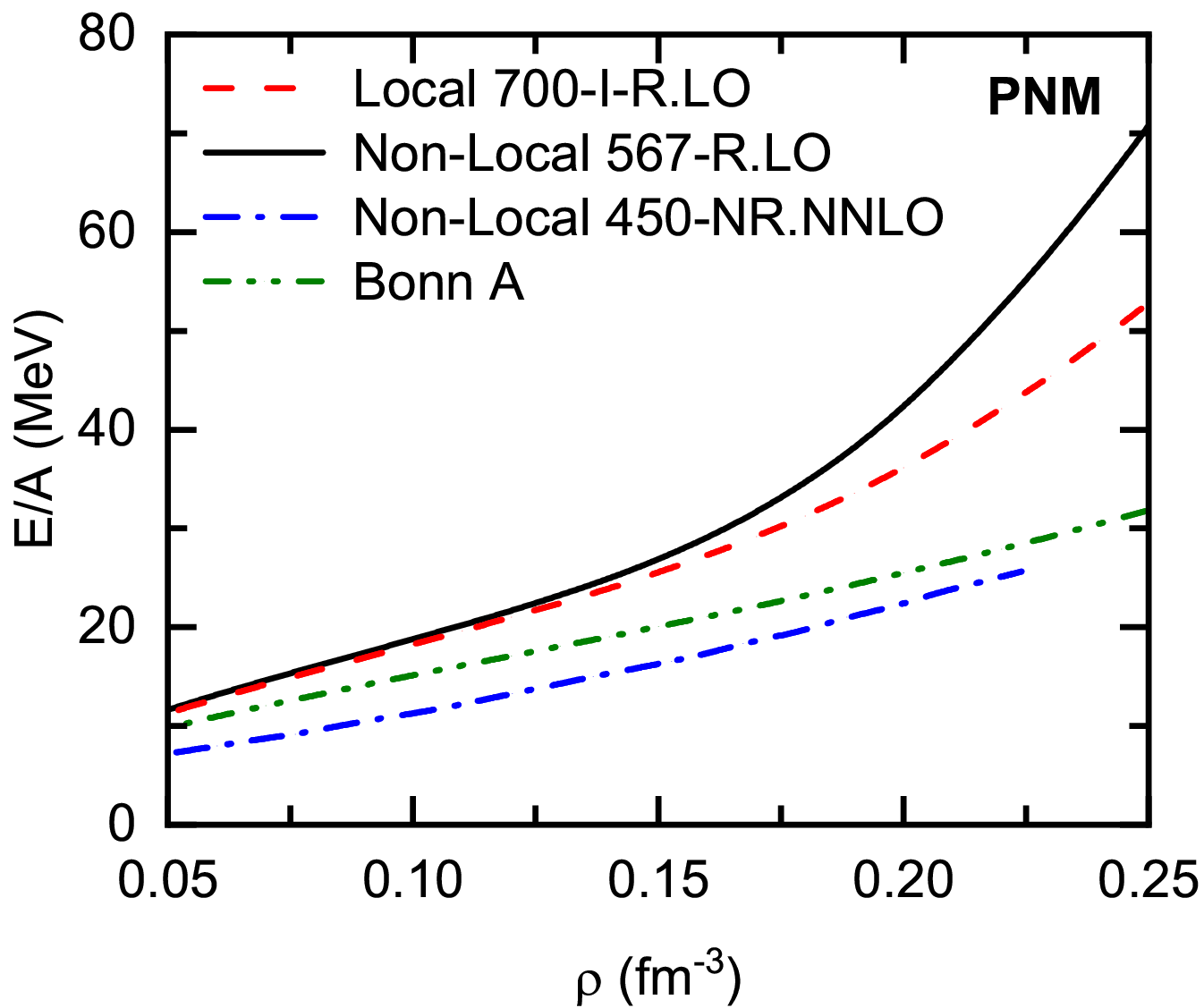}
\caption{\label{fig3}~(Color online)~Energy per nucleon~($E/A$)~of PNM as a function of the density~$\rho$, obtained with the LO covariant chiral interactions with local regularization function~(red dash line), in comparison with the previous results with a nonlocal regularization function~\cite{Zou:2023quo}(solid line), nonrelativistic NNLO chiral nuclear force~\cite{Sammarruca:2021mhv}~(blue dash-dot line) and Bonn A potential~(green dash-dot-dot line). The numbers in the legend indicate the momentum cutoff values.}
\end{figure}

In Fig.~\ref{fig3}, we compare the EoSs for PNM calculated using four different nuclear forces: the LO covariant chiral interactions with local and nonlocal regularization functions, the nonrelativistic NNLO chiral NN + 3NFs, and the Bonn A potential. Compared to the results from the Bonn A potential, the LO covariant chiral interactions with both local and nonlocal regularization functions yield stiffer EoSs for PNM. 
This discrepancy arises because the LO covariant chiral interactions overestimate the repulsive contributions in P-wave, as indicated by the inadequate description of the P-wave phase shifts~\cite{Ren:2016jna,Zou:2023quo}.
Incorporating higher-order contributions in the chiral expansion is expected to mitigate this issue.

\section{Summary and Outlook}
In summary, we have investigated the equation of state of symmetric nuclear matter and pure neutron matter using the RBHF theory with the LO covariant chiral interactions.
Different from the previous studies of NN phase shifts and nuclear matter, the covariant chiral interactions are regulated by local functions.
We found that the saturation properties of symmetric nuclear matter can be reasonably described in a certain cutoff range, similar to the previous RBHF study using nonlocal LO covariant chiral interactions. 
Therefore, the present results provide the validation of the local LO covariant chiral interactions in nuclear matter. The present results provide confidence to apply the LO covariant chiral interactions with local regularization for finite nuclei.   

In future work, 
we will implement the LO covariant chiral interactions with local regularization functions to the investigations of finite nuclei, using the RBHF~\cite{Shen:2016bva,Shen2017PRC} or the relativistic quantum Monte Carlo~\cite{Yang2022PLB,Yang:2024wsg} methods.
In addition,
it would be interesting to perform studies of symmetric nuclear matter and pure neutron matter  with higher-order covariant chiral interactions and examine the convergence of chiral expansion.

\section*{Acknowledgments}
This work was supported in part by the National Natural Science Foundation of China under Grants No.12435006, No.12435007, No.12475117, No.12141501,  No.123B2080, the National Key R\&D Program of China under Grant No.2024YFE0109803, and the National Key Laboratory of Neutron Science and Technology NST202401016.

\bibliography{apssamp}

\newpage

\appendix

\end{document}